\documentclass[12pt]{article}
\usepackage{graphicx}
\usepackage{epstopdf}
\usepackage{amsmath}
 \numberwithin{equation}{subsection}

\newcommand\pubnumber{DPF2013-42}
\newcommand\pubdate{\today}

\def\napoli{School of Physics and Astronomy\\
University of Minnesota, Minneapolis, MN 55455, USA}
\def\support{\footnote{On behalf of the NO$\nu$A collaboration, email: bianjm@physics.umn.edu}}

\def\Title#1{\begin{center} {\Large #1 } \end{center}}
\def\Author#1{\begin{center}{ \sc #1} \end{center}}
\def\Address#1{\begin{center}{ \it #1} \end{center}}

\newcommand\pubblock{\rightline{\begin{tabular}{l} \pubnumber\\
         \pubdate  \end{tabular}}}
\newenvironment{Abstract}{\begin{quotation}  }{\end{quotation}}
\newenvironment{Presented}{\begin{quotation} \begin{center} 
             PRESENTED AT\end{center}\bigskip 
      \begin{center}\begin{large}}{\end{large}\end{center} \end{quotation}}

\def\beq{\begin{equation}}
\def\eeq#1{\label{#1}\end{equation}}
\def\eeqn{\end{equation}}

\def\beqa{\begin{eqnarray}}
\def\eeqa#1{\label{#1}\end{eqnarray}}
\def\eeqan{\end{eqnarray}}

\let\bar=\overbar

\def\Dslash{\not{\hbox{\kern-4pt $D$}}}
\def\dslash{\not{\hbox{\kern-2pt $\del$}}}

\def\msb{{\bar{\ssstyle M \kern -1pt S}}}

\begin{document}
\begin{titlepage}
\pubblock

\vfill
\Title{The NO$\nu$A experiment: overview and status}
\vfill
\Author{ Jianming Bian\support}
\Address{\napoli}
\vfill
\begin{Abstract}
The NO$\nu$A experiment is a long-baseline accelerator based neutrino oscillation experiment. It uses the upgraded Fermilab NuMI beam and measures electron-neutrino appearance and muon-neutrino disappearance at its far detector in Ash River, Minnesota. Goals of the experiment include measurements of $\theta_{13}$, the neutrino-mass hierarchy and the CP-violating phase.  NO$\nu$A has begun to take data this year and will have its first physics results in 2014.  This talk provides an overview of the scientific reach of the NO$\nu$A experiment, the status of detector construction and physics analysis and a first glimpse of far-detector data.
\end{Abstract}
\vfill
\begin{Presented}
DPF 2013\\
The Meeting of the American Physical Society\\
Division of Particles and Fields\\
Santa Cruz, California, August 13--17, 2013\\
\end{Presented}
\vfill
\end{titlepage}
\def\thefootnote{\fnsymbol{footnote}}
\setcounter{footnote}{0}

\section{Introduction}
\begin{figure}[htb]
\centering
\includegraphics[height=8cm]{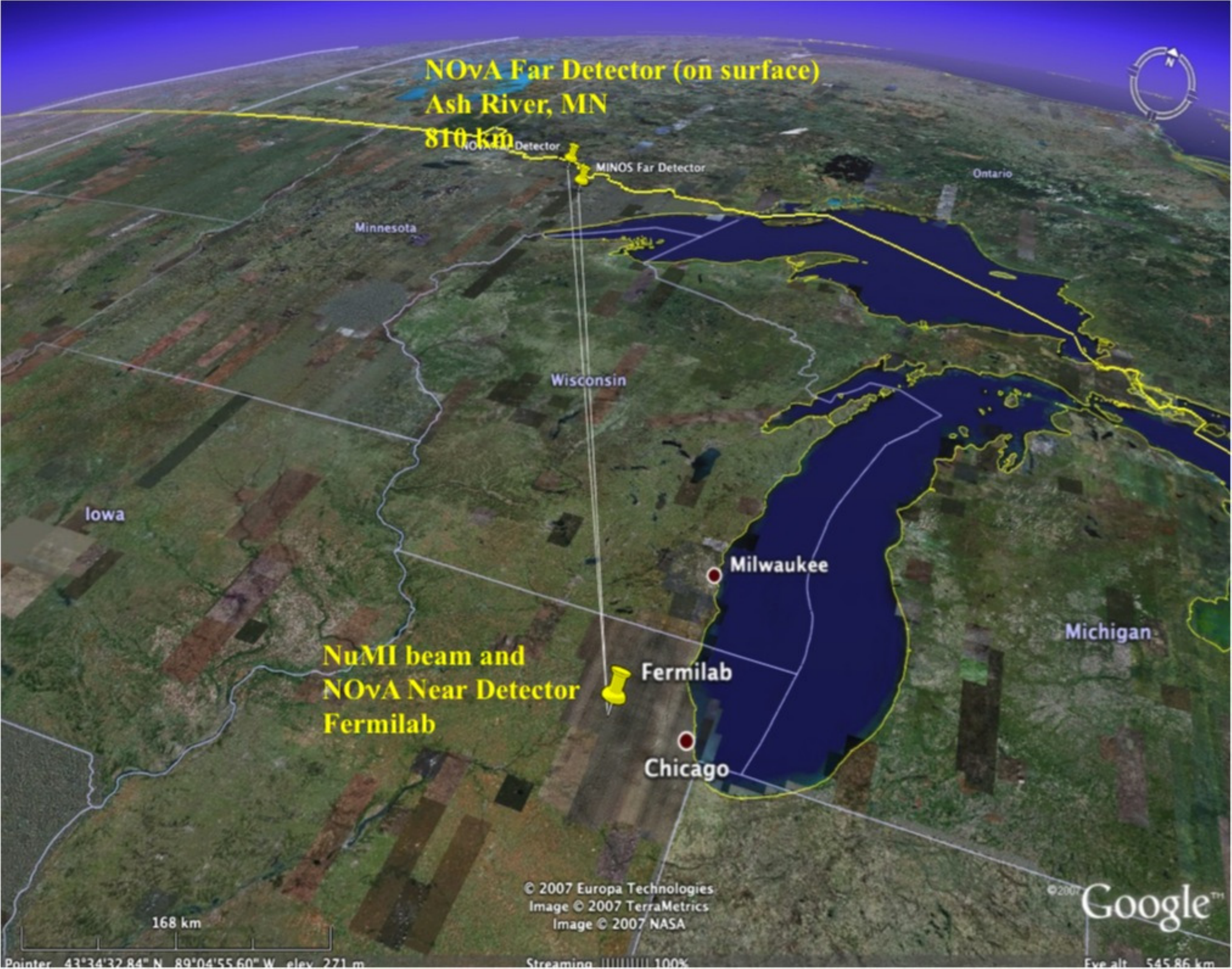}
\caption{The NO$\nu$A experiment}
\label{fig:beam}
\end{figure}

\begin{figure}[htb]
\centering
\includegraphics[height=6cm]{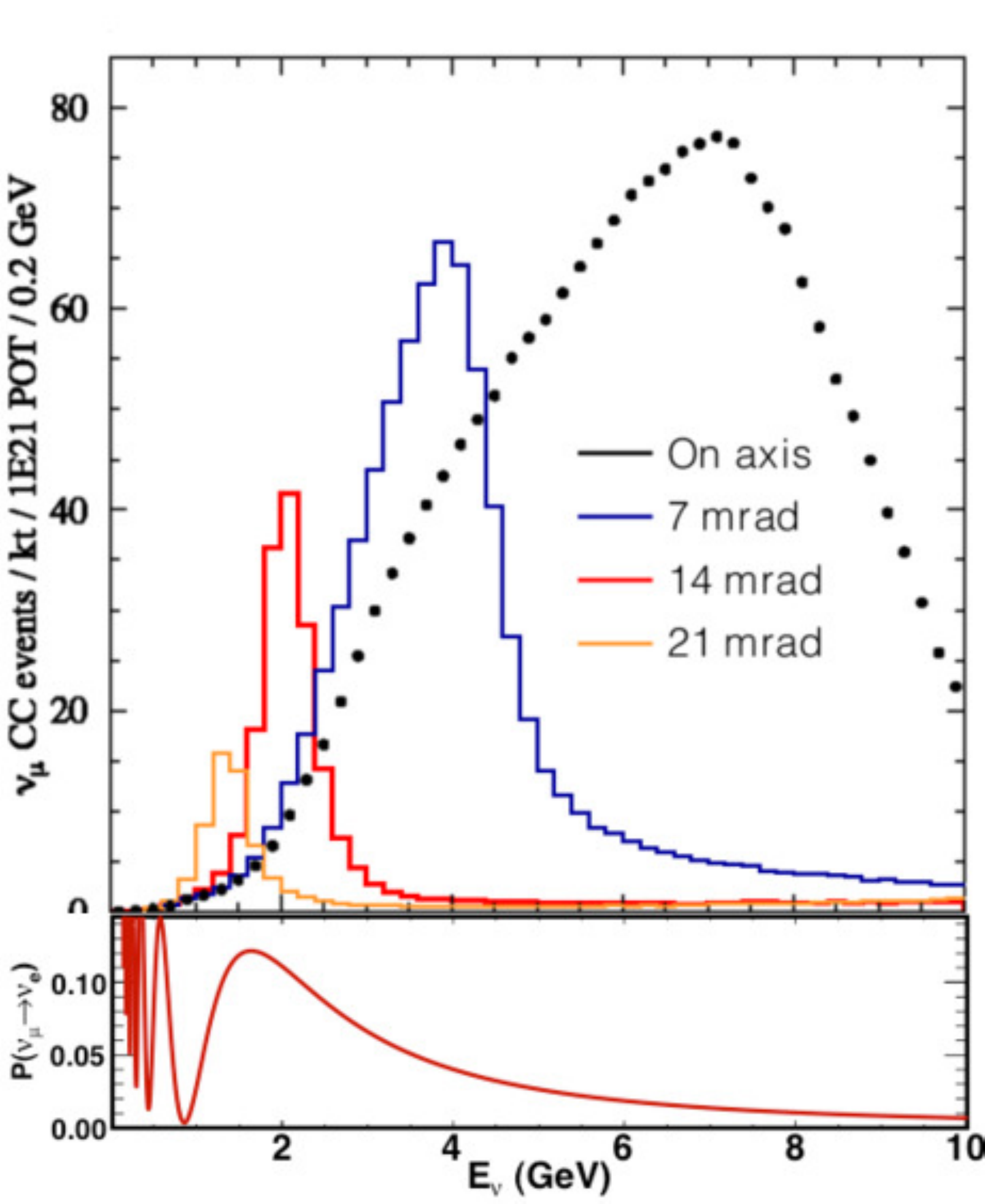}
\caption{Simulated neutrino energy spectra for $\nu_{\mu}$ charged current
interactions in detectors sited 0, 7, 14, and 21 mrad off the NuMI
beam axis. NO$\nu$A uses 14 mrad. The probability of $\nu_\mu\to\nu_e$ as a function of energy are shown in the bottom.}
\label{fig:expectedspectrum}
\end{figure}

NuMI Off-Axis $\nu_e$ Appearance Experiment (NO$\nu$A) is a 2-detector neutrino oscillation experiment which is optimized for $\nu_e$ identification. NO$\nu$A uses the NuMI muon neutrino beam at Fermilab as its neutrino source~\cite{ref:numi}, and a 14 kt liquid scintillator far detector at a distance of 810 km (Ash river, Minnesota) to detect the oscillated beam. It has the longest possible baseline, for the NuMI beamline in the United States, which maximizes the matter effect and allows a measurement of the neutrino mass ordering . NO$\nu$A equips a  $\sim$300 ton, functionally identical near detector located at Fermilab to measure unoscillated beam neutrinos to estimate backgrounds in the far detector. A map for NO$\nu$A beam and detectors is shown in Figure~\ref{fig:beam}.
The far detector is sited 14 mrad off-axis to produce a narrow-band beam around the oscillation maximum region ($\sim$2 GeV),  as shown in Figure~\ref{fig:expectedspectrum}. Combining with  final state interaction cross sections, the off-axis setting provides NO$\nu$A an optimized signal/background separation. 

NO$\nu$A measures $\nu_e$ ($\bar{\nu_e}$) appearance probability and $\nu_\mu$ ($\bar{\nu_\mu}$) disappearance probability with neutrino and anti-neutrino beams. The $\nu_e$ ($\bar{\nu_e}$) appearance experiment investigates (1) the neutrino masses hierarchy (the mass ordering of $\nu_3$ and the other two), (2) the CP violation phase in the neutrino sector, (3) PMNS mixing angle $\theta_{13}$, and (4) $\theta_{23}$ octant (whether $\theta_{23}>$ or $ <45^\circ$). For the $\nu_\mu$ ($\bar{\nu_\mu}$) disappearance experiment, NO$\nu$A will perform  precise measurement on the atmospheric oscillation parameters $|\Delta m_{32}^2|$ and $\theta_{23}$. Because NO$\nu$A has large detectors, it can also be used to study other physics such as neutrino cross sections, neutrino magnetic moment, supernova, monopoles, sterile neutrinos, and non-standard neutrino interactions.

The NO$\nu$A detectors are fine grained and highly active tracking calorimeters. They consist of plastic (PVC) extrusions filled with liquid-scintillator, with wavelength shifting fibers (WLS) connected to avalanche photodiodes (APDs).  The cross-sectional size of detector cells are about 6 cm $\times$ 4 cm. Each cell extends the full width or height of the detector, 15.6 m in the FD and 4.1 m in the ND. Extrusions are assembled in alternating layers of vertical and horizontal extrusions, so 3-D hit information is available for tracking. The 14-kton Far Detector has 344,064 cells and the 0.3-kton Near Detector has 18,000 cells. Each plane (cell width) of the detectors is  just 0.15 radiation lengths ($X_0$). This level of granularity helps greatly to separate electrons from $\pi^0$ backgrounds. 

 Each detector cell has a wavelength-shifting fiber connected to an Avalanche
 Photodiode (APD). Scintillation light emitted isotropically is captured
 in wavelength-shifting fibers that convert the wavelength to the APDs' sensitive region. APDs have two substantial advantages over other photodetectors: high quantum efficiency and uniform spectral quantum efficiency. The high APD quantum efficiency enables the use of very long scintillator modules, thus significantly reducing the electronics channel count.

\section{Construction status and cosmic ray data}

The construction of NO$\nu$A  is in good shape. The site at Ash River
for the Far Detector was completed in 2012 and the Far Detector's
assembly and commissioning is in full swing. By July 22, 2013, $64\%$ of
blocks have been installed, $41\%$ of the detector has been filled and
$15\%$ of blocks have been outfitted with electronics. The completion of
the Far Detecter is expected by May, 2014. The construction of the Near
Detector is also initiated: cavern excavation  for the Near Detector is
complete and the muon catcher was installed Aug. 1, 2013. The first half
of the detector will be installed by the end of 2014 and the second half
will be complete by summer 2014. 

\begin{figure}[htb]
\centering
\includegraphics[height=6cm]{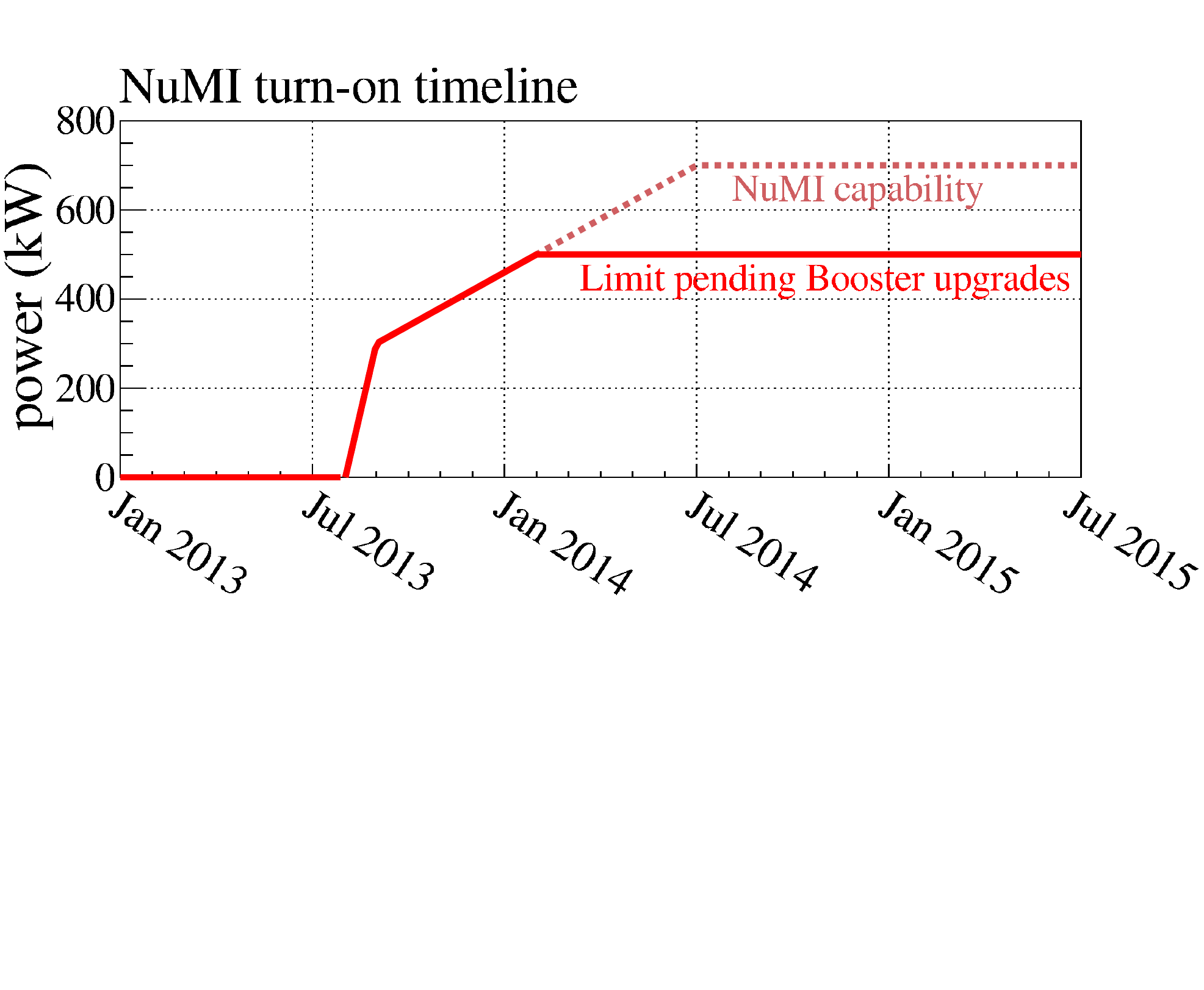}
\caption{The sketch of the evolution of NuMI power in early running after the 2012/2013 upgrades.}
\label{fig:beamass}
\end{figure}

The NuMI beam is undergoing upgrades to increase its power from 350 kW
to 700 kW. Fermilab has completed a series of upgrades to the Main
Injector and Recycler Rings to reduce the cycle time from 2.2 s to 1.3 s
which improves the intensity from 300 to 700 kW. The neutrino beamline
has also been optimized for NO$\nu$A. Commissioning of accelerators is
underway. Routine operation of the neutrino beam starts in September 2013, with a beam power of 3.6 kW. The beam intensity will ramp to 500 kW in 2013 and later to 700 kW. The sketch of the evolution of NuMI power in early running after the 2012/2013 upgrades can be found in Figure~\ref{fig:beamass}.

\begin{figure}
\centering
\begin{minipage}{.33\textwidth}
  \centering
  \includegraphics[width=4 cm]{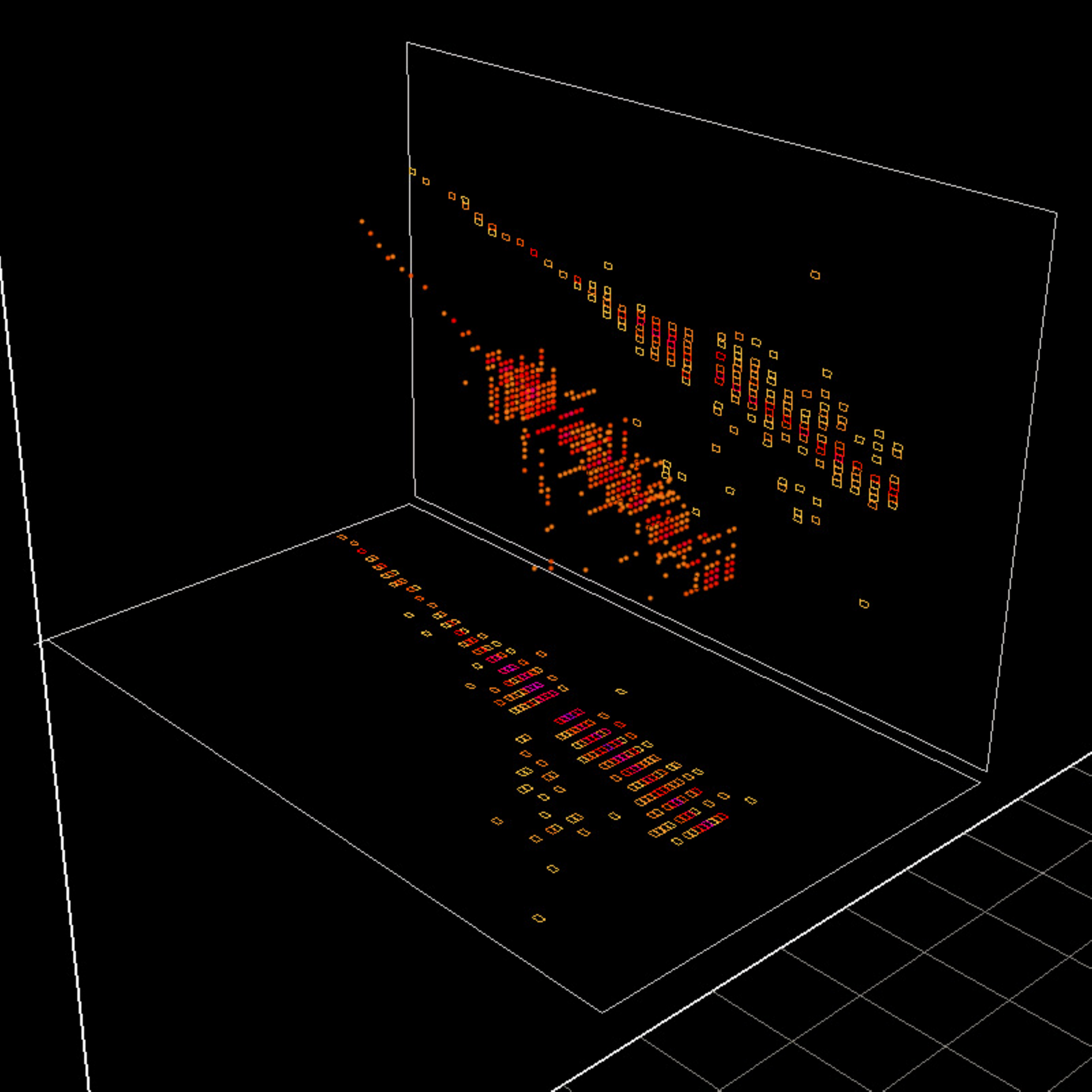}
\end{minipage}%
\begin{minipage}{.33\textwidth}
  \centering
  \includegraphics[width=6 cm]{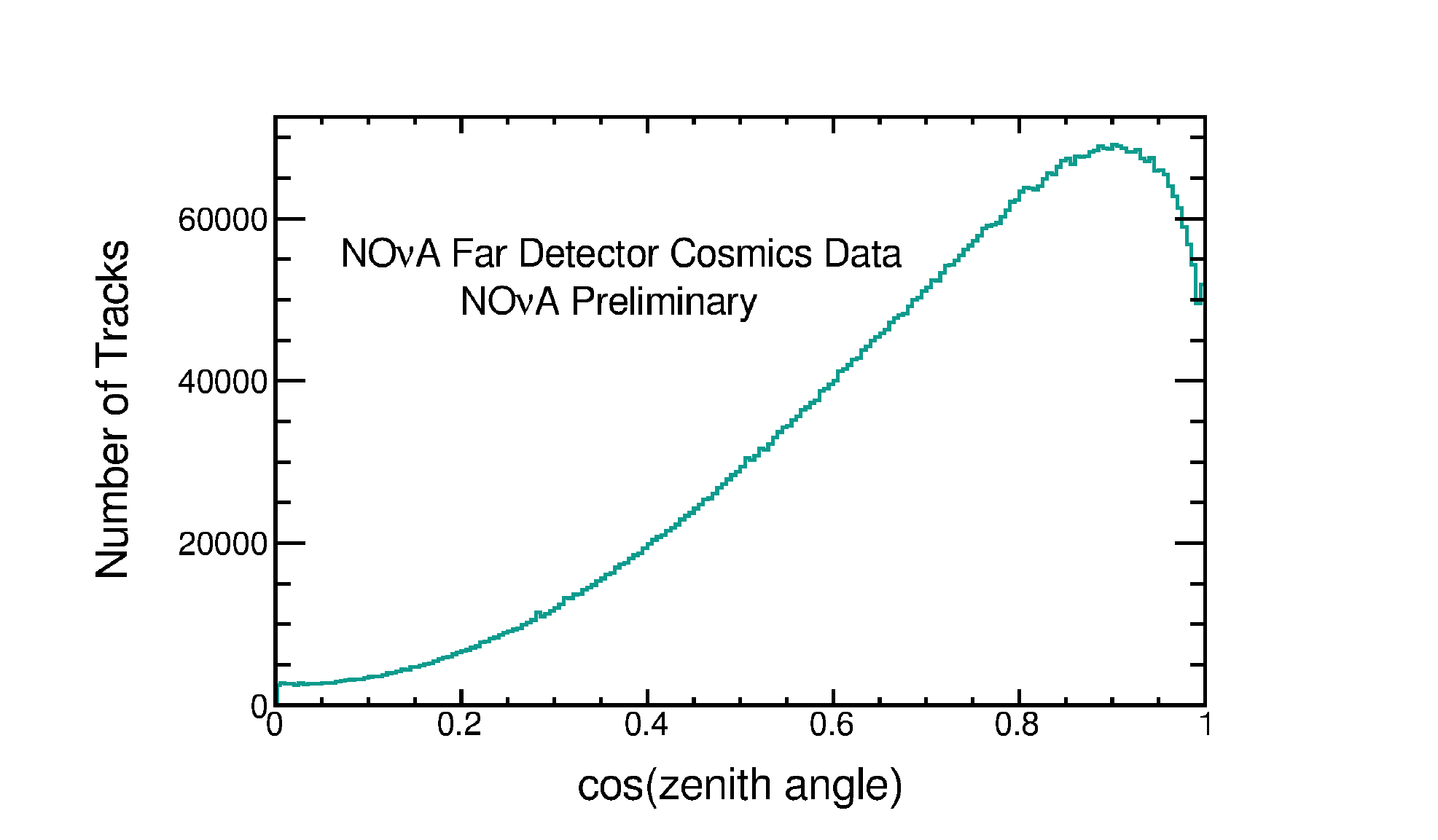}
\end{minipage}
\begin{minipage}{.33\textwidth}
  \centering
  \includegraphics[width=6 cm]{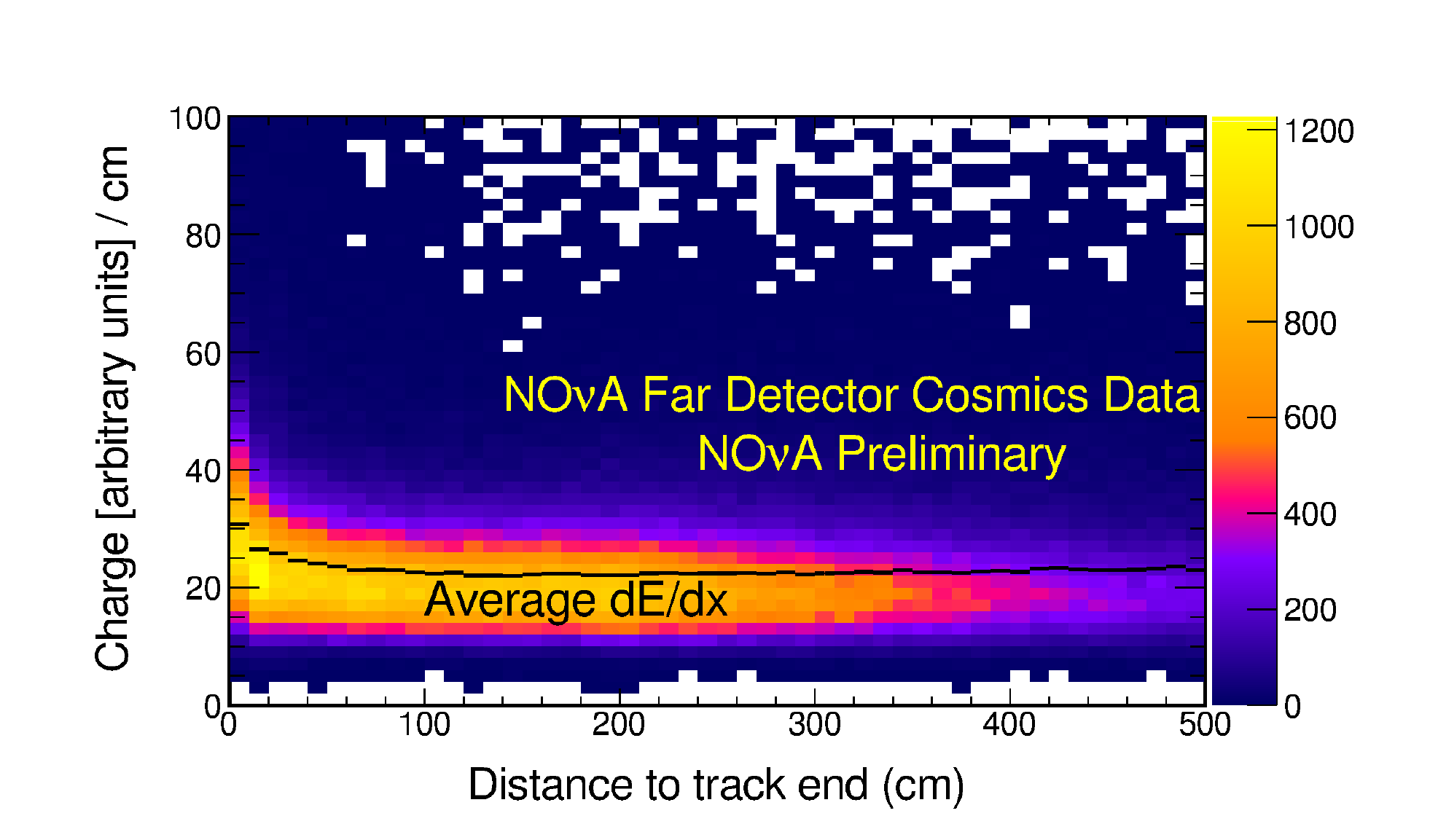}
\end{minipage}
  \caption{The cosmic ray data taken and reconstructed by NO$\nu$A, with
  the first kton of the Far Detector which has been instrumented since May 21, 2013. left: event display for a cosmic ray event, middle: zenith angle distribution for cosmic ray data, right: stop muon d$E/$d$x$ with respect to distance from track end in cosmic ray data}
  \label{fig:cos}
\end{figure}

With the first kton of the Far Detector, instrumented May 21, 2013,
NO$\nu$A has started to take cosmic ray data. Reconstruction and calibration algorithms are already tested on these data, as shown in Figure~  \ref{fig:cos}.

\section{Physics reach and analysis preparation}

The $\nu_e$ appearance analysis at NO$\nu$A will use $18\times10^{20}$ POT (protons on target) for neutrino running and $18\times10^{20}$ POT for anti-neutrino running, which can be accumulated in six years at design intensities. The probability of $\nu_\mu(\bar{\nu_\mu})\to\nu_e (\bar{\nu_e})$ for NO$\nu$A can be written as:

\begin{equation}\label{prob}
\begin{split}
P(\nu_\mu\to\nu_e)  \approx & \sin^22\theta_{13}\sin^2\theta_{23}\frac{\sin^2(A-1)\Delta}{(A-1)^2} \\
                                                   & +2\alpha\sin\theta_{13}\cos\delta\sin2\theta_{12}\sin2\theta_{23}\frac{\sin A\Delta}{A}\frac{\sin(A-1)\Delta}{(A-1)}\cos\Delta\\
                                                   & -2\alpha\sin\theta_{13}\sin\delta\sin2\theta_{12}\sin2\theta_{23}\frac{\sin A\Delta}{A}\frac{\sin(A-1)\Delta}{(A-1)}\sin\Delta\\
{\rm where} ~~ \Delta \equiv & \frac{\Delta m^2_{31}L}{4E},\\
                                   A \equiv & \pm\frac{G_f n_e L}{\sqrt{2}\Delta}\approx\frac{E}{11 GeV}~('+' for~\nu_\mu\to\nu_e~~and~~'-'~for~\bar{\nu_\mu}\to\bar{\nu_e}).
\end{split}
\end{equation}

All three terms of this equation include $\theta_{13}$. Because recent
results by reactor neutrino experiments demonstrate $\theta_{13}$ is
large, NO$\nu$A will be able to collect a substantial number of events for  $\nu_\mu(\bar{\nu_\mu})\to\nu_e (\bar{\nu_e})$. As  shown in the 2nd and 3rd terms of the equation, the probability is enhanced or suppressed due to matter effects. The sign of $A$ in the matter effect is determined by neutrino vs. anti-neutrino running and the amplitude of the enhancement or suppression depends on $\theta_{13}$, mass hierarchy ($\Delta m_{31}^2$), CP violation phase ($\delta$), and $\theta_{23}$. So by measuring probabilities of $\nu_\mu\to\nu_e $ and $\bar{\nu_\mu}\to\bar{\nu_e}$, we could be able to solve the mass hierarchy and provide information on $\delta_{CP}$ simultaneously. 

\begin{figure}[htb]
\centering
\includegraphics[height=6cm]{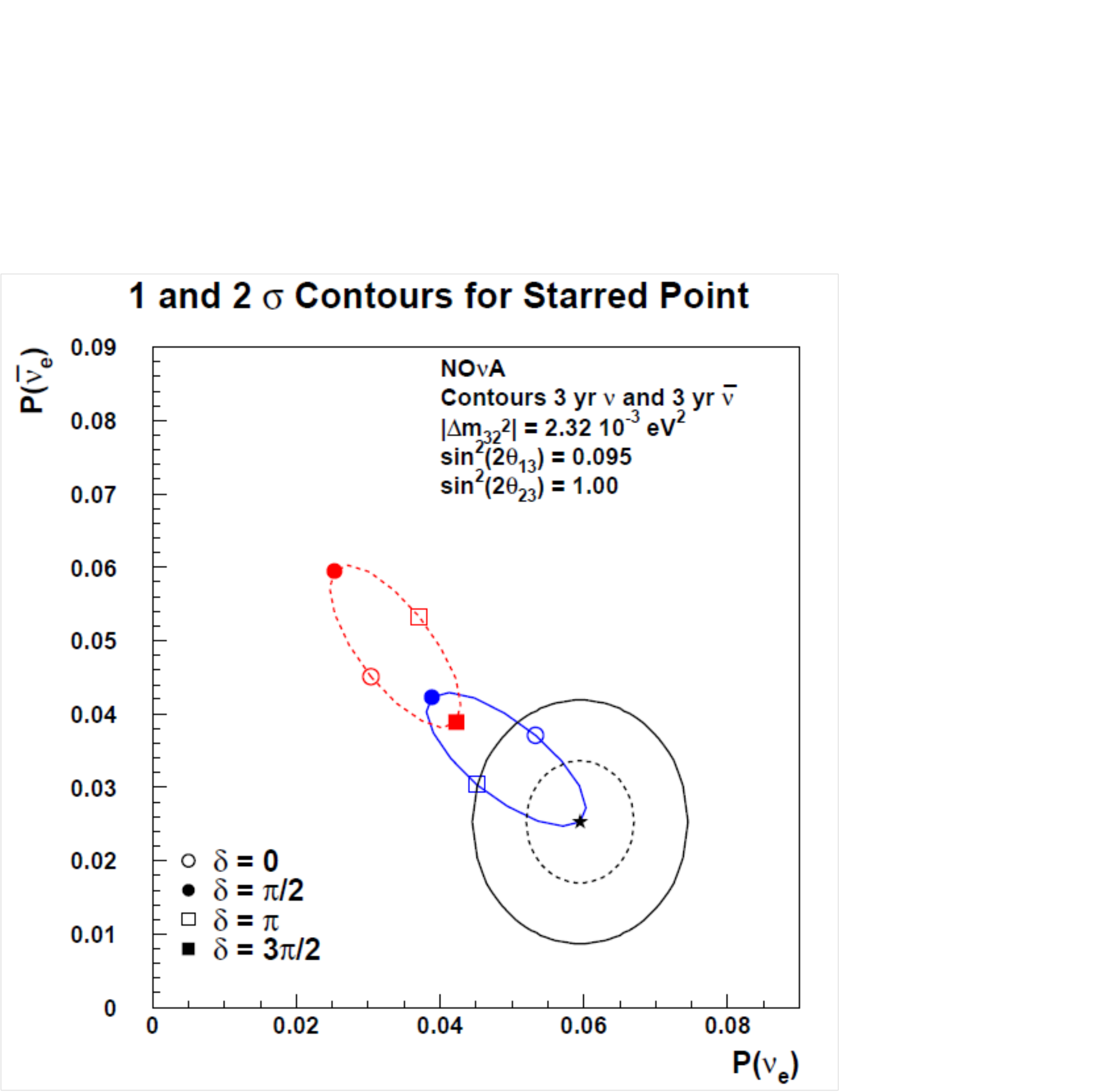}
\caption{The principle of the NO$\nu$A $\nu_e$ appearance measurements.
All possible values for the 2-GeV appearance probabilities $\nu_\mu\to\nu_e $ and $\bar{\nu_\mu}\to\bar{\nu_e}$ are shown. The solid
blue (dashed red) ellipse corresponds to the normal (inverse) hierarchy scenario, with $\delta$ varying as one moves around each ellipse. $1\sigma$ and $2\sigma$ sensitivities are shown in solid and dashed black for the test case at $\delta=3\pi/2$ , normal hierarchy (starred point).}\label{fig:biprob}
\end{figure}

Figure~\ref{fig:biprob} demonstrates the principle by which NO$\nu$A
determines the mass hierarchy and measures the CP phase for a given set
of oscillation parameters. The probability of $\nu_\mu\to\nu_e $ is
represented by a point on the x-axis and $\bar{\nu_\mu}\to\bar{\nu_e}$
is represented by a point on the y-axis. One can find that
$P(\nu_\mu\to\nu_e)$  vs.  $P(\bar{\nu_\mu}\to\bar{\nu_e})$ appears to
be two ellipses -  the blue ellipse is the trace for the assumption of
normal mass hierarchy and red is for the assumption of inverse mass
hierarchy. On each ellipse the position is determined by CP phase. As an
example, if $\delta=3\pi/2$, NO$\nu$A can make a measurement at the starred point. According to the $1\sigma$ (dashed line) and $2\sigma$ (solid line) contours for the six-year of NO$\nu$A running, all inverted hierarchy scenarios are excluded at $>2\sigma$. The precision of probabilities measurement depends on $\theta_{13}$, while a large $\theta_{13}$ also reduces the overlap area of normal hierarchy and inverse hierarchy ellipses, so it is very good news for NO$\nu$A that $\theta_{13}$ is not small.

The preparation for NO$\nu$A physics analysis is in good shape. The reconstruction chain for cell, track, shower and vertex is well developed and shows good performance. To identify $\nu_e$ from various backgrounds, we have developed three particle identification algorithms which can meet with the physics requirement: (1) ANN, an artificial neural network using shower shape based likelihood for particle hypotheses, (2) LEM, a technique that matches events to a Monte Carlo library, and (3) RVP, a boosted decision tree based on simple reconstructed quantities. We also have tools to determine backgrounds in the near detector and extrapolate them to the far detector.

\begin{figure}[htb]
\centering
\includegraphics[height=5cm]{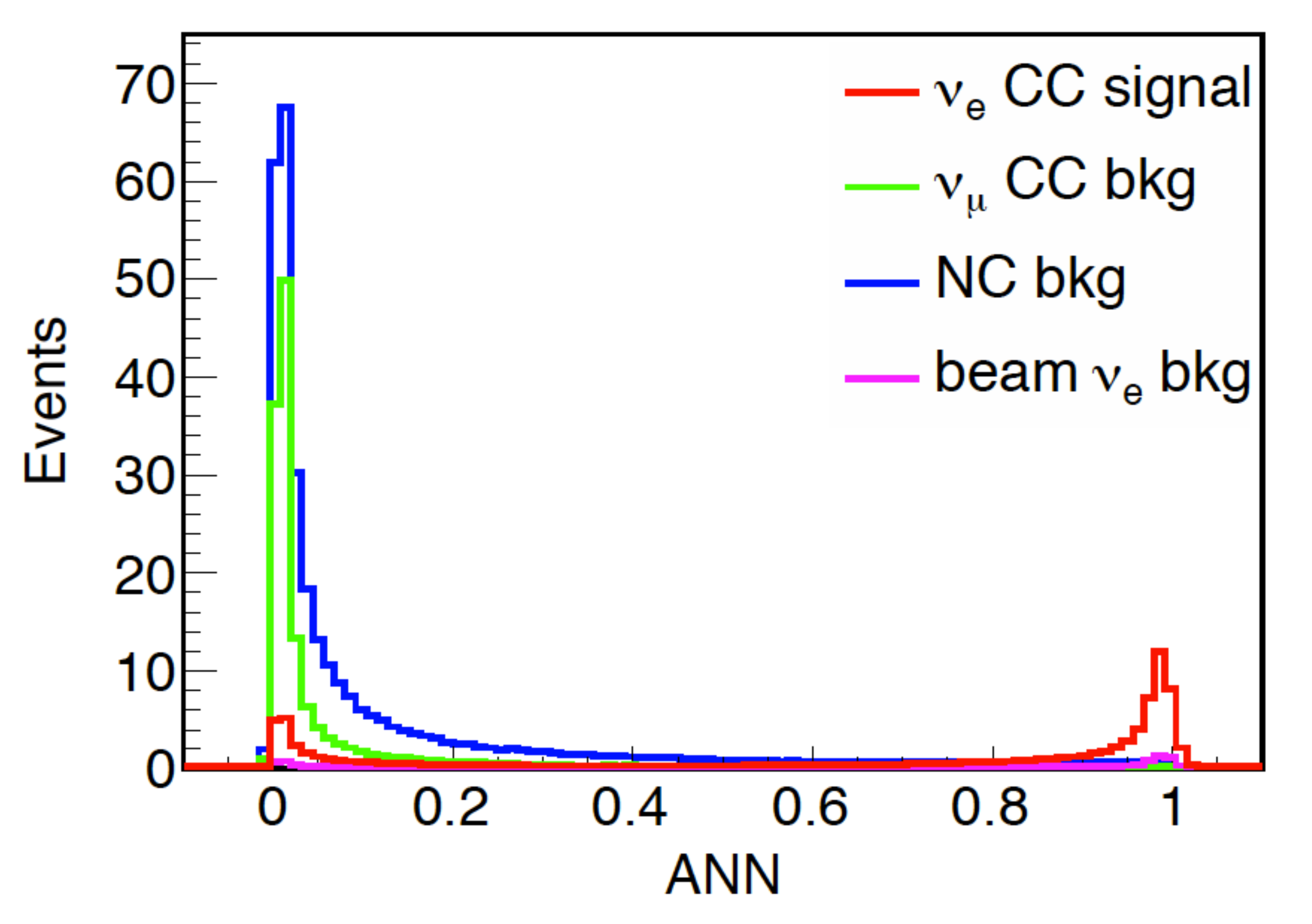}
\caption{The neural network output for $\nu_e$ identification at NO$\nu$A.}
\label{fig:ann}
\end{figure}

We take the ANN method as an example to explain the $\nu_e$
identification at NO$\nu$A. The basic idea is to use the shower energy profile to distinguish electron from $\mu$, $\pi^0/\gamma$ and other hadrons.  Different particles have very different energy deposit behaviors in the detector. For example, the electron ionizes in the first few planes then starts a shower, the photon is a shower with a gap in the first few planes,  and the muon deposits a long minimum ionizing particle (MIP) track. This makes it possible to identify particles by comparison of shower shapes with different particle hypotheses.  Because the NO$\nu$A detectors have fine cells and can provide excellent energy resolution and tracking, we can perform this comparison plane by plane for the longitudinal direction and cell by cell for transverse direction. In this way we can make use of all energy profile information in a shower.  Using these likelihoods, a neural network has been trained and applied to identification of electrons for analysis of candidate $\nu_e$ events, as shown in Figure~\ref{fig:ann}. 

\begin{figure}
\centering
\begin{minipage}{.5\textwidth}
  \centering
  \includegraphics[width=5 cm]{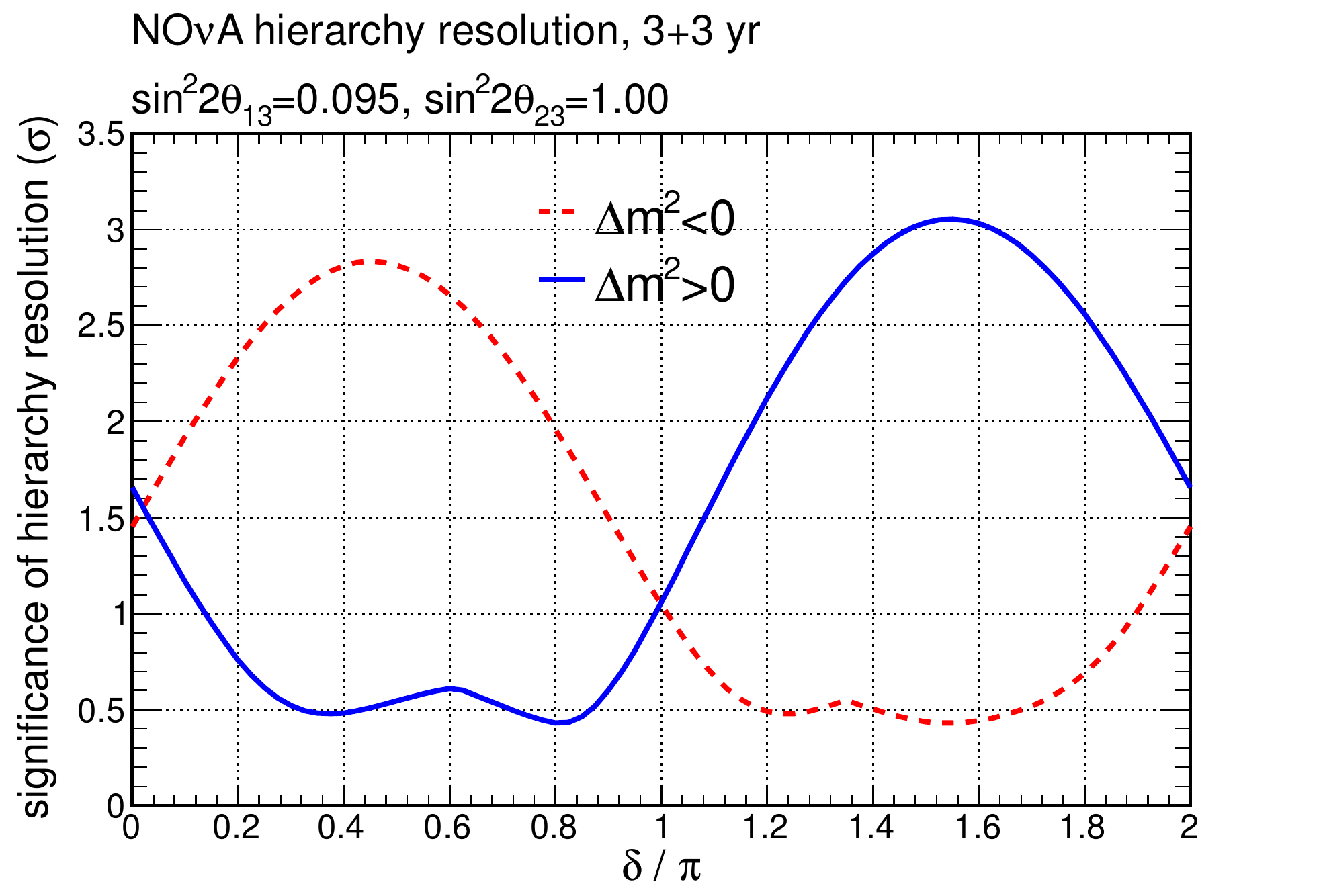}
\end{minipage}%
\begin{minipage}{.5\textwidth}
  \centering
  \includegraphics[width=5 cm]{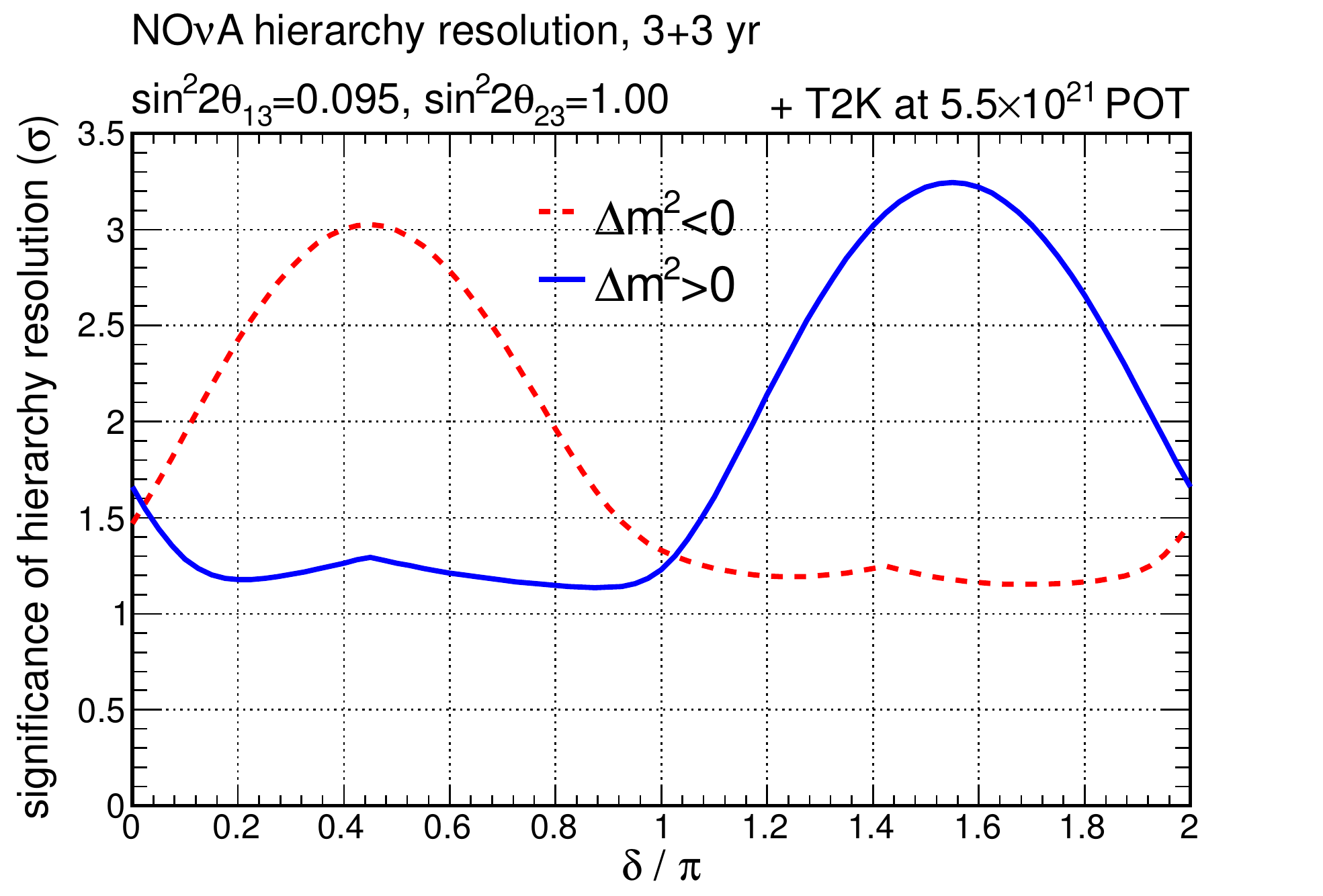}
\end{minipage}
  \caption{The significance of mass hierarchy resolution for:  NO$\nu$A (left) and NO$\nu$A+T2K (right), as a function of the true value of $\delta$.}
  \label{fig:mh}
\end{figure}

\begin{figure}
\centering
\begin{minipage}{.5\textwidth}
  \centering
  \includegraphics[width=5 cm]{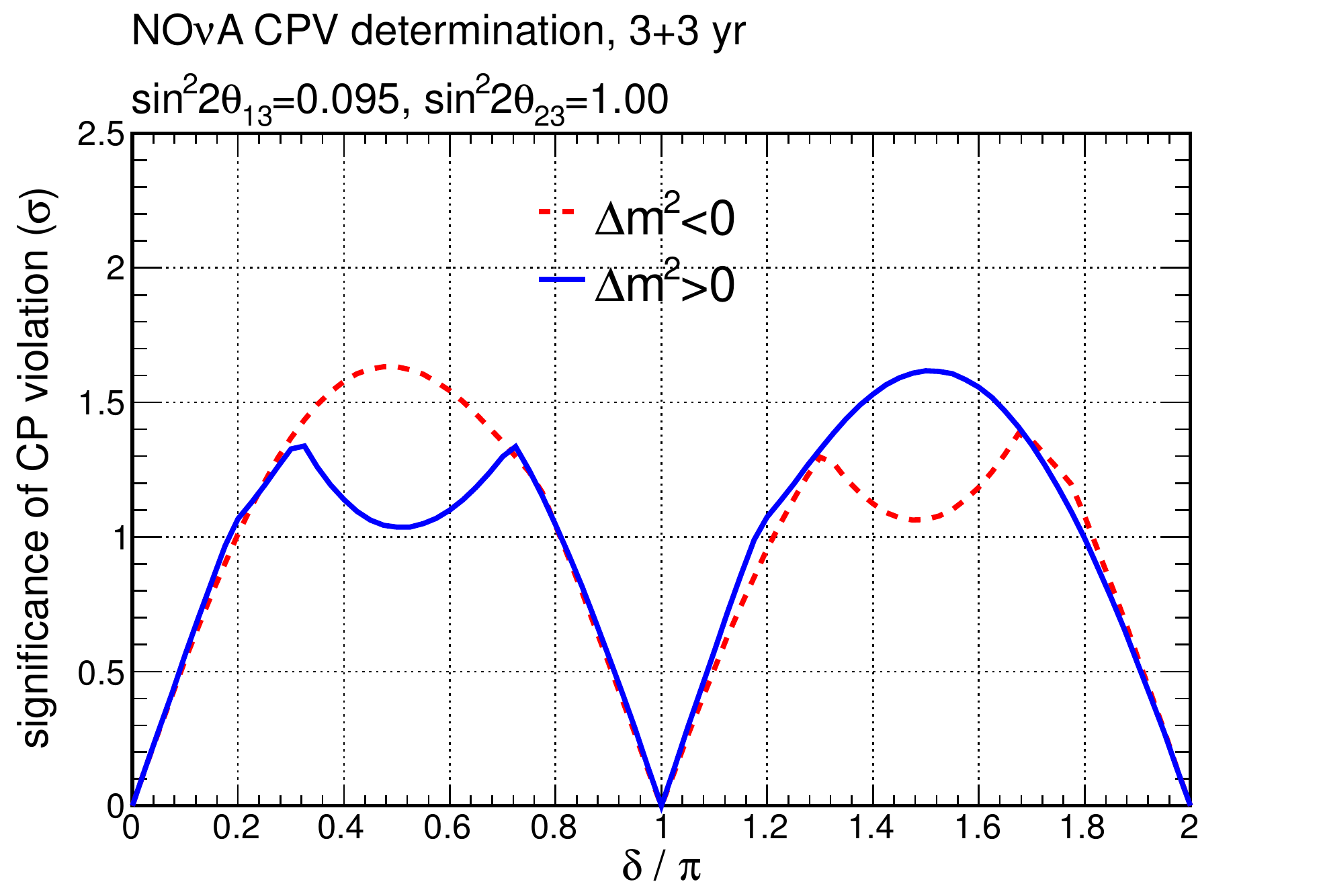}
\end{minipage}%
\begin{minipage}{.5\textwidth}
  \centering
  \includegraphics[width=5 cm]{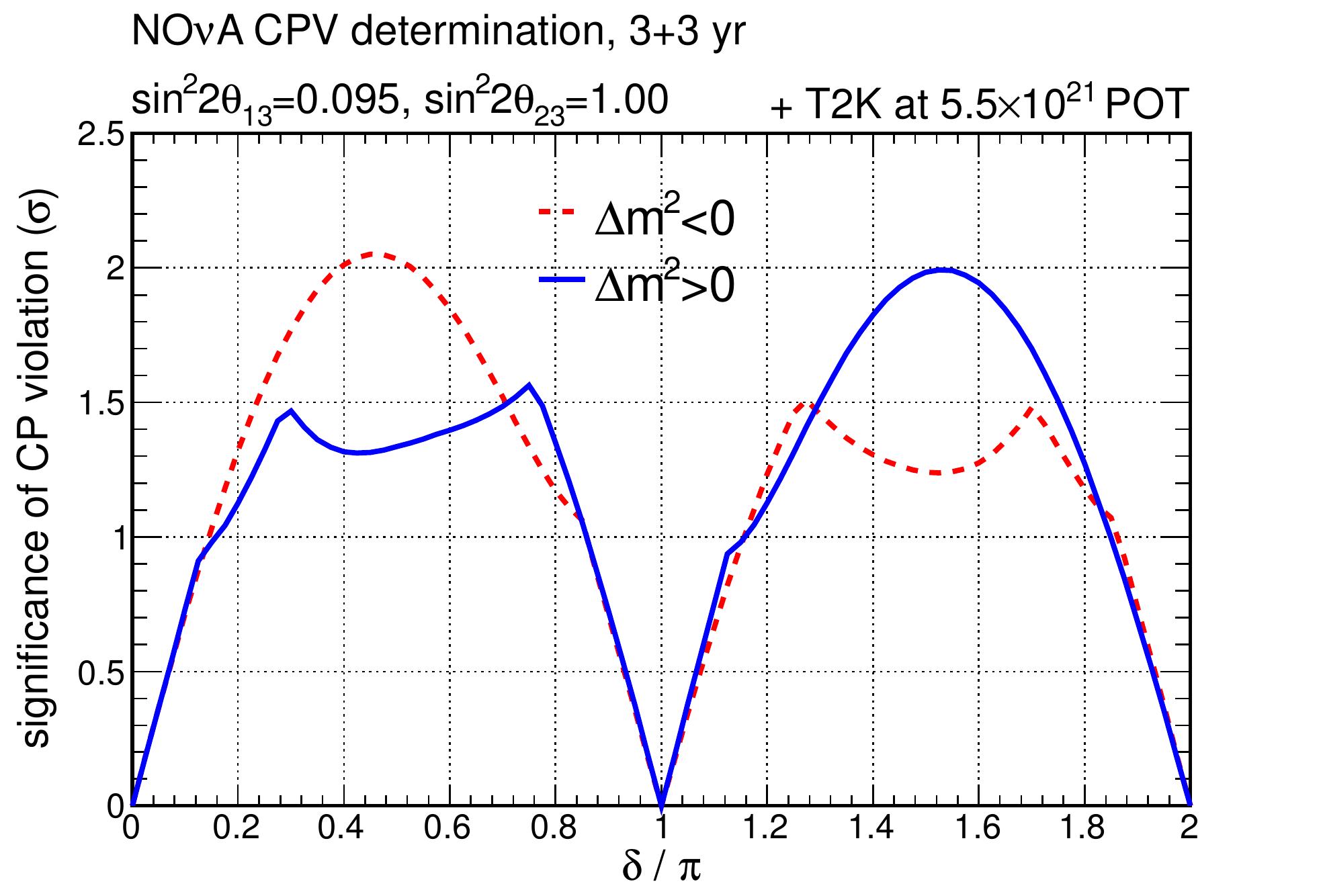}
\end{minipage}
  \caption{The significance of CP violation for:  NO$\nu$A (left) and NO$\nu$A+T2K (right), as a function of the true value of $\delta$.}
  \label{fig:cpv}
\end{figure}

Using the Far Detector Monte Carlo sample, the $\nu_e$ appearance
analysis has been performed with full simulation, reconstruction,
selection, and analysis framework. The resulting significance of the
mass hierarchy is shown explicitly as a function of $\delta$ in
Figure~\ref{fig:mh} (left). One can find that in about $40\%$ of the range of $\delta$, NO$\nu$A will determine the mass hierarchy better than $2\sigma$. Combining with T2K's result, potential degeneracies in the NO$\nu$A measurement can be lifted to exceed $1\sigma$, as shown in Figure~\ref{fig:mh} (right). This analysis also provides some sensitivity for $\delta$, as shown in Figure~\ref{fig:cpv}. 

\begin{figure}
\centering
\begin{minipage}{.5\textwidth}
  \centering
\includegraphics[height=5cm]{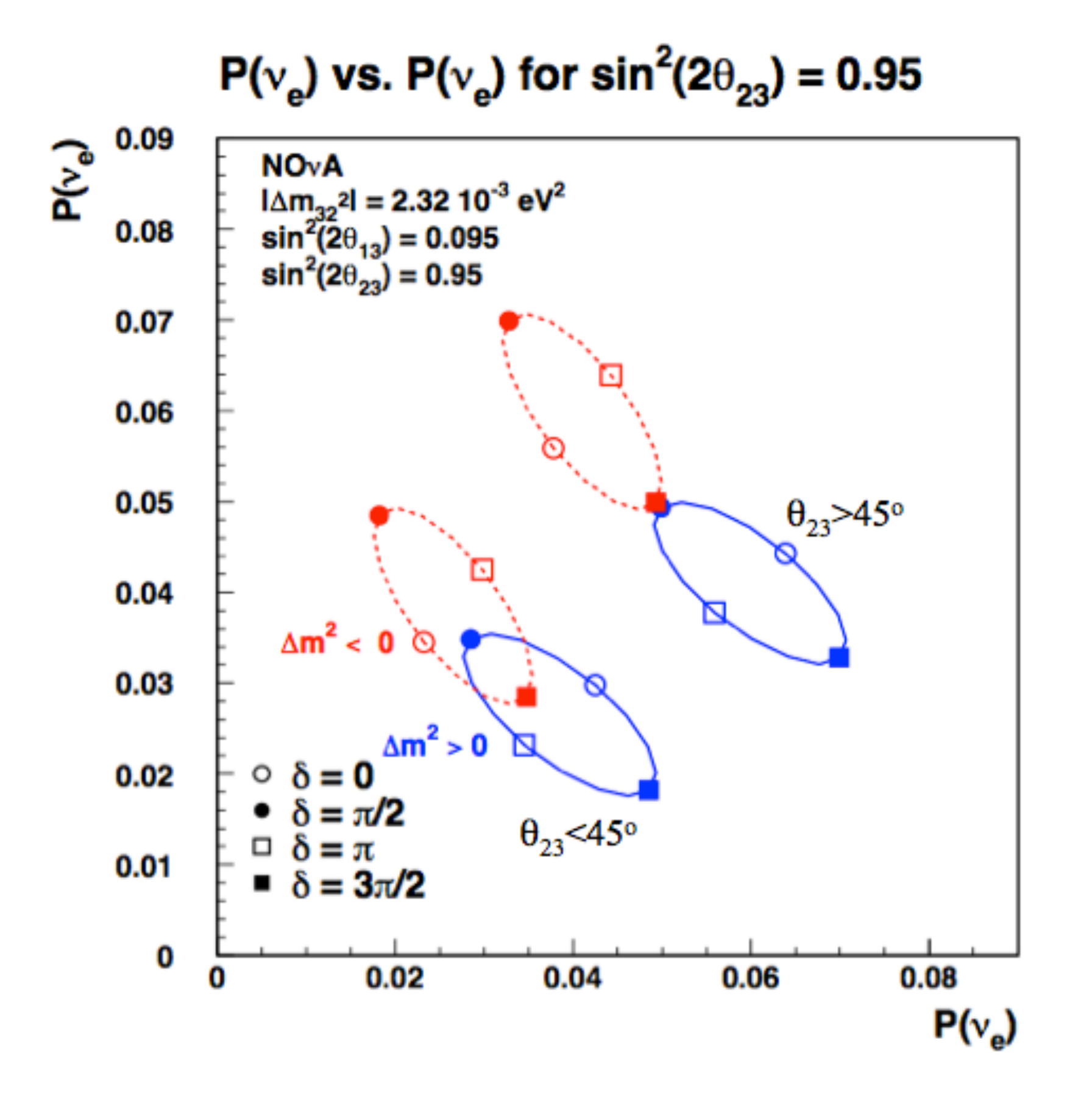}
\end{minipage}%
\begin{minipage}{.5\textwidth}
  \centering
\includegraphics[height=4cm]{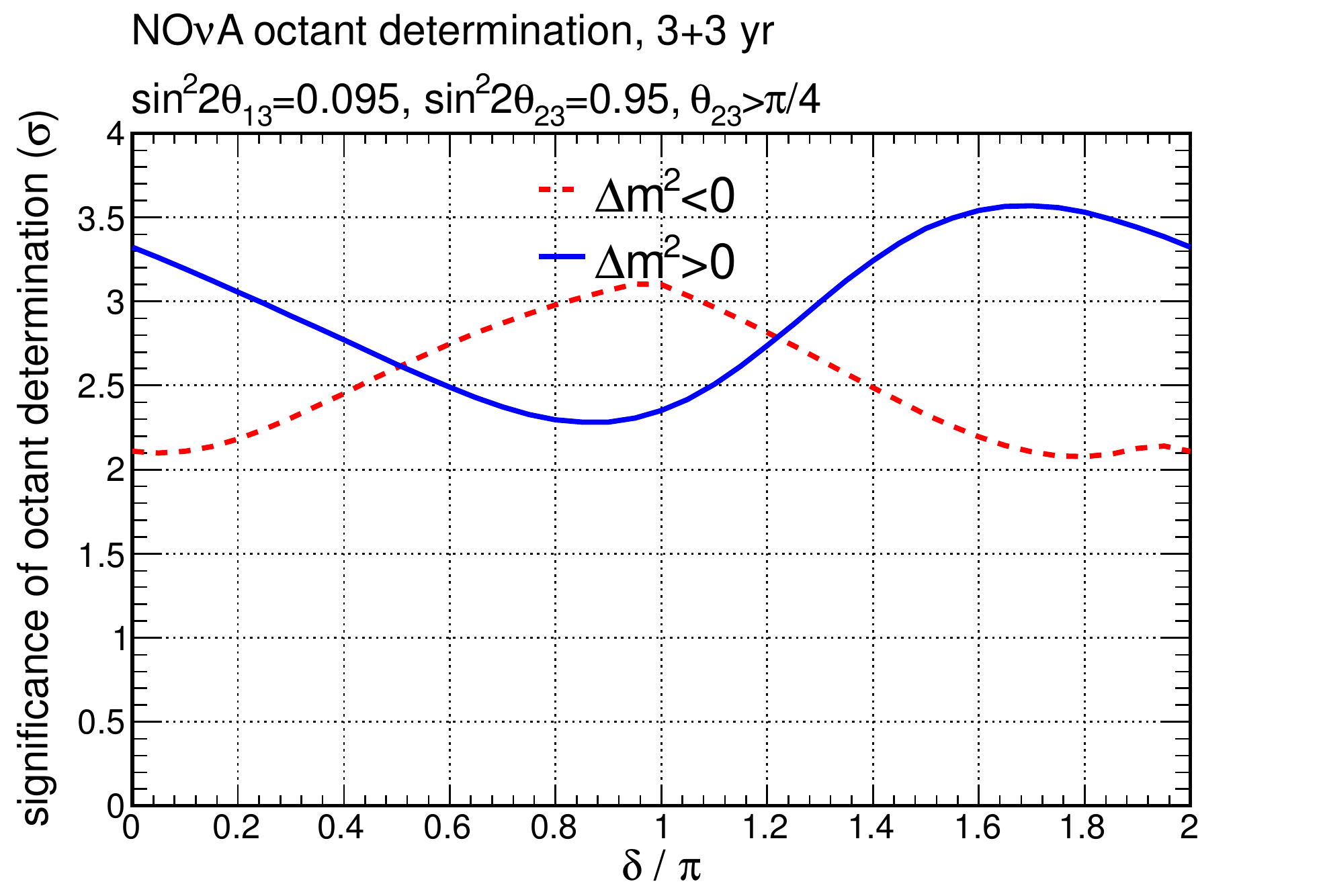}
\end{minipage}
\caption{left: The idea for the $\theta_{23}$ octant measurement, with an assumption that $\sin^2 2\theta_{23} = 0.95$ and $\theta_{23}>45^{\circ}$, right: NO$\nu$A's sensitivity to determine the $\theta_{23}$ octant as a function of the true value of delta.} 
  \label{fig:octantsens}
\end{figure}

This example to extract the mass hierarchy and $\delta$ is under the assumption that $\sin^2(2\theta_{23}) = 1$. If $\sin^2(2\theta_{23})$ is not maximal there is an ambiguity as to whether $\theta_{23}$ is larger or smaller than $45^\circ$, i.e., the octant of $\theta_{23}$. The $\sin^2(\theta_{23})$ term is crucial in comparing accelerator to reactor experiments. Because $P(\nu_\mu\to\nu_e)$ is in proportion to $\sin^2(\theta_{23})\sin^2(2\theta_{13})$, it can be used to determine the $\theta_{23}$ octant. Figure~\ref{fig:octantsens} (left) shows how non-maximal $\sin^2(2\theta_{23}) $ influences the NO$\nu$A $\nu_e$ appearance measurements. As marked in the figure,  the probability ellipses are in different areas when $\theta_{23}$ $>$ or $<$ $45^\circ$. This dependence makes NO$\nu$A able to  determine the $\theta_{23}$ octant with mass hierarchy and $\delta$ simultaneously. Figure~\ref{fig:octantsens} (right) shows NO$\nu$A's sensitivity to determine the $\theta_{23}$ octant as a function of the true value of delta, for each of the possible hierarchies. One can find that for the assumption $\sin^2 2\theta_{23} = 0.95$ and $\theta_{23}>45^{\circ}$, NO$\nu$A will determine the $\theta_{23}$ octant better than $2\sigma$.

\section{Summary}

NO$\nu$A has the best chance to investigate mass hierarchy. It can
determine the $\theta_{23}$  octant and will provide first information
on CP violation. NO$\nu$A has started to take cosmic ray data and the
NuMI beam has returned in September, 2013. NO$\nu$A detector construction is in very good shape and will complete in 2014. Many software tools are in place for physics analysis.  We are focusing on commissioning our far detector and working towards first physics results in summer 2014.

\end{document}